\documentclass{iopart}
\usepackage{graphicx}

\begin{document}
\title{Ab Initio Liquid Hydrogen Muon Cooling Simulations with ELMS in ICOOL}
\author{W. W. M. Allison, J. H. Cobb, S. J. Holmes, Physics Dept., Oxford\\
 R. C. Fernow, R. B. Palmer, BNL, Upton, Long Island, New York}

\begin{abstract}
This paper presents new theoretical results on the passage of muons through liquid hydrogen which have been 
confirmed in a recent experiment. These are used to demonstrate that muon bunches may be compressed by ionisation 
cooling more effectively than suggested by previous calculations. 

Muon cooling depends on the differential cross section for energy loss and scattering of muons. We have calculated 
this cross section for liquid H$_2$ from first principles and atomic data, avoiding traditional assumptions. Thence, 
2-D probability maps of energy loss and scattering in mm-scale thicknesses are derived by folding, and stored in a 
database. Large first-order correlations between energy loss and scattering are found for H$_2$, which are absent in 
other simulations. This code is named ELMS, Energy Loss \& Multiple Scattering. Single particle trajectories may 
then be tracked by Monte Carlo sampling from this database on a scale of 1 mm or less. This processor has been 
inserted into the cooling code ICOOL. Significant improvements in 6-D muon cooling are predicted compared with 
previous predictions based on GEANT. This is examined in various geometries. The large correlation effect is found 
to have only a small effect on cooling. The experimental scattering observed for liquid H$_2$ in the MUSCAT 
experiment has recently been reported to be in good agreement with the ELMS prediction, but in poor agreement with 
GEANT simulation.
\end{abstract}

\section{Introduction}
New particle accelerators may be built that are based on intense beams of high energy muons. Such beams would 
provide a source of highly collimated neutrinos, and later muon colliding beams~\cite{mc1,mc2}. To achieve such 
beams the muons must be compressed and cooled to increase the phase space density. It has been pointed out that 
passage of the beam through low $Z$ material, followed by RF acceleration to replace the longitudinal momentum lost, 
would cool the beam transversely while suffering the least heating due to scattering. A rigourous understanding of 
the passage of muons through liquid hydrogen is therefore required. The longitudinal phase space dynamics of the 
muons depends on both the muon velocity and the change in velocity; a thorough understanding and accurate simulation 
is therefore required in this case also.\footnote{A fuller exposition of the theory and simulation will be published 
later.}

\section{Simulation process}
\subsection{Cross section}
Each collision of a muon in a material involves both energy transfer and momentum transfer. For example to predict 
reliably the cooling and heating effects of the passage of a muon bunch with energies $E$ through a thickness of 
liquid hydrogen requires a knowledge of the appropriate cross section for energy transfer $\cal E$ and transverse 
momentum transfer $p_t$, \[\frac{{\rm d}^2\sigma}{{\rm d}{\cal E}{\rm d}p_t}(E, {\cal E}, p_t).\]

This cross section is determined uniquely by: \begin{itemize} \item the known $\mu$-proton cross 
section\footnote{This is the Rosenbluth modification of the Rutherford cross section.} with $Q^2$ 
dependence\footnote{The variable $Q^2$ is the 4-momentum transfer squared, $p^2-({\cal E}/c)^2$, where $p$ is the 
3-momentum transfer.} modified by the screening of the proton charge by the known electron atomic wavefunction; 
\item the excitation and ionisation cross section of molecular hydrogen that can be deduced~\cite{WJ80} from the 
known photoabsorption spectrum of H$_2$~\cite{JB02}; \item the $\mu$-electron cross section for $Q^2$ greater than 
that involved in photoelectric atomic ionisation.\footnote{This is the Dirac modification of the Rutherford cross 
section.}\end{itemize}
 
As discussed elsewhere~\cite{WJ80, W03, SJH06}, these parts do not interfere and parts 2 \& 3 are joined by the 
Bethe sumrule~\cite{WJ80}. The H$_2$ photoabsorption spectrum requires a small correction for density. This is well 
determined by the measured refractive index of liquid H$_2$. The only uncertainty is the role played by higher 
multipoles at high $Q^2$. Reasonable modifications suggest that this effect is everywhere less than 
2\%~\cite{SJH06}.
 
The cross section, simplified in standard treatments into separate energy loss and scattering processes, involves 
correlations between energy loss and scattering. This is important in hydrogen where scattering as well as energy 
loss through collisions with atomic electrons is significant. 

The cross section includes the full range of excitation, ionisation, multiple scattering, relativistic recoil, 
Cherenkov effect, delta rays, nuclear and atomic form factors in a unified way. Bremsstrahlung contributions have 
been calculated but are insignificant here. 
 
\subsection{Folding probabilities for thin-absorber database}

The collision frequency in liquid H$_2$ is several million per m, while some significant collisions are rare. This 
relates to the $1/Q^4$ of the Rutherford cross section and the large range of $Q^2$ available. Therefore tracking 
particles by Monte Carlo at the microscopic level is statistically poorly behaved. 

A different approach is to use the cross section to tabulate 2-D probability maps of energy and scattering by 
folding the maps from thinner samples. Thus we have constructed such maps for thicknesses of a mm or less for 138 
momenta on a logarithmic scale from 5 to 53000 MeV/c, close enough to permit interpolation to 1\% accuracy. The maps 
are on a 5\% logarithmic grid in energy loss and $p_t$. In folding together maps of thinner samples the azimuth of 
the scattering in the two layers is stepped through. The layers have to be thin enough that it may be assumed that 
the cross section and the path traversed within the layer are not changed as a result of collisions within the 
layer. The result is a database of normalised probability maps which may be sampled.

Derived secondary formulae such as Moli\`{e}re scattering, Bethe--Bloch mean energy loss, multiple scattering from 
the radiation length and other short cuts that make Gaussian assumptions are not used. The calculations are not 
tuned as there are no free parameters.

\subsection{MC sampling of thin absorber database for general cooling studies}

For absorbers of arbitary thickness the database may be sampled and individual muons tracked taking full account of 
changes in direction and momentum, following each step. The mm-scale step length is decreased with $\beta^2$ and is 
affected by interpolation. Tracking is fast.

The ELMS sampling algorithm with its database has been embedded in the ICOOL~\cite{icoolref} muon tracking code, 
taking care to maintain step-length compatibility.\footnote{Standard ICOOL uses algorithms very close to those of 
GEANT3. For materials other than hydrogen, e.g. absorber windows, ELMS-ICOOL uses standard energy loss and 
scattering algorithms.}

\section{Results}

\subsection{Correlation effects}

\begin{figure}[tb]
\centering
\includegraphics[width=105mm]{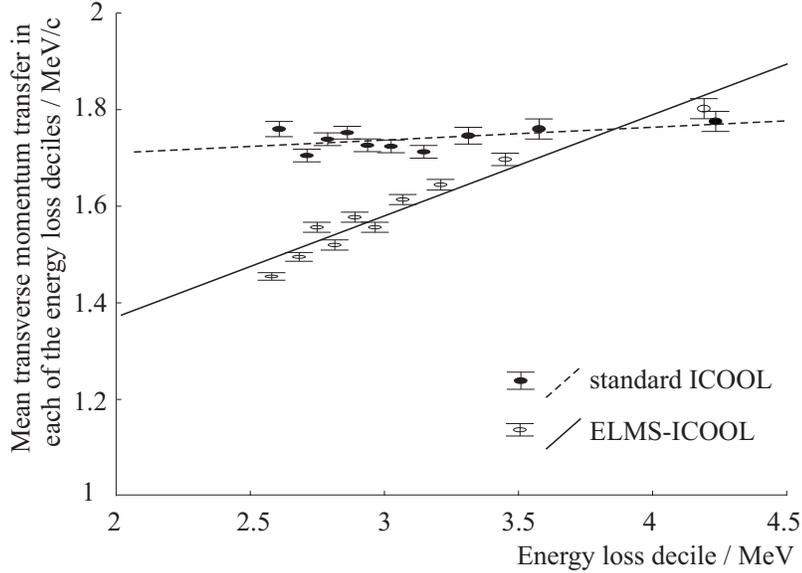}
\caption{Correlation between scattering and energy loss at 200MeV/c in 10cm H$_2$ as simulated by standard ICOOL and 
by ELMS-ICOOL. The mean transverse momentum transfer is plotted for each energy loss decile.\label{f:slabcor}}
\end{figure}

Fig.~\ref{f:slabcor} shows that correlation between scattering and energy loss persists even in finite absorber 
thicknesses. A second order effect whereby a muon that loses a large energy then becomes more liable to suffer large 
scattering is evident in the standard ICOOL simulation. However the first order effect due to the differential cross 
section included in ELMS-ICOOL is much larger. We have studied the effect of suppressing this correlation in 
ELMS-ICOOL simulations, and found it has no effect on cooling characteristics~\cite{SJH06}. 

\subsection{Comparison with the MUSCAT experiment at 172MeV/c}

\begin{figure}[tb]
\centering
\includegraphics[width=70mm]{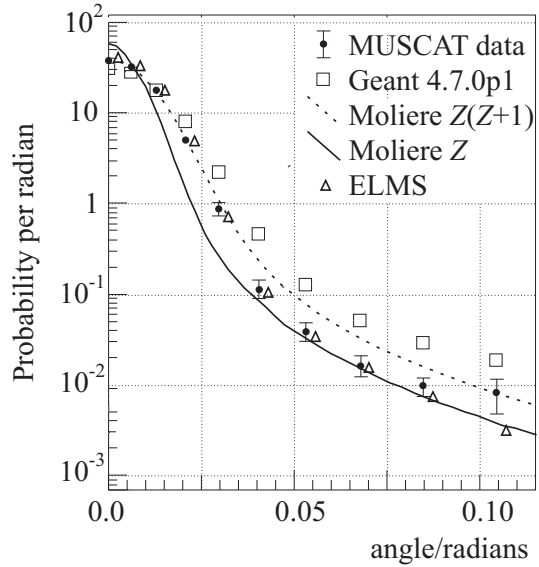}
\caption{Experimental distribution in multiple scattering in 15.9cm of liquid H$_2$ from the MUSCAT 
collaboration~\cite{musc}, compared with various calculations including ELMS. The data are shown with error bars 
where these are significant. The ELMS points are all displaced to the right of the data points for 
clarity.\label{f:muscfig22}}
\end{figure}

Multiple scattering distributions have been compared with the recent observations of the MUSCAT 
collaboration~\cite{musc}. These data corrected for the effect of windows are shown for 15.9cm of liquid H$_2$ in 
Fig.~\ref{f:muscfig22}. The agreement with ELMS is remarkable. The Moli\`{e}re calculation when it includes electron 
scattering fits the data at small angles while at large angles the limited $\mu$-e centre-of-mass momentum prevents 
such a contribution. This was pointed out by Tollestrup~\cite{toll}. Such kinematic effects are naturally included 
in ELMS. GEANT~\cite{geantref} gives a poor description. 

For cooling the energy loss is as important as the scattering distribution. However there is less difference between 
calculations. For example, the mean range as a function of momentum determined by ELMS agrees with tabulated values 
of range~\cite{SJH06}. This is a non-trivial check.

We may conclude that ELMS is a good basis on which to predict the performance of a muon cooler. 

Similar calculations for other materials have not yet been made. However differences are not expected to be so 
marked since the combination of energy loss and scattering from atomic electrons is less significant for greater 
values of $Z$. 

\subsection{Simulation of MICE at 200MeV/c}

We have considered the cooling in transverse emittance by a repeated MICE\footnote{Muon Ionisation Cooling 
Experiment~\cite{mice, SJH06}} channel 100m long including the effect of windows. Simulations with standard ICOOL 
and with ELMS-ICOOL showed that 90\% of the beam survived, equilibrium emittance was not reached, and transverse 
cooling of 55\% (standard ICOOL) and 59\% (ELMS-ICOOL) were predicted. 

\subsection{Simulation of RFOFO ring at 200MeV/c}

\begin{figure*}
\centering
\includegraphics[width=130mm]{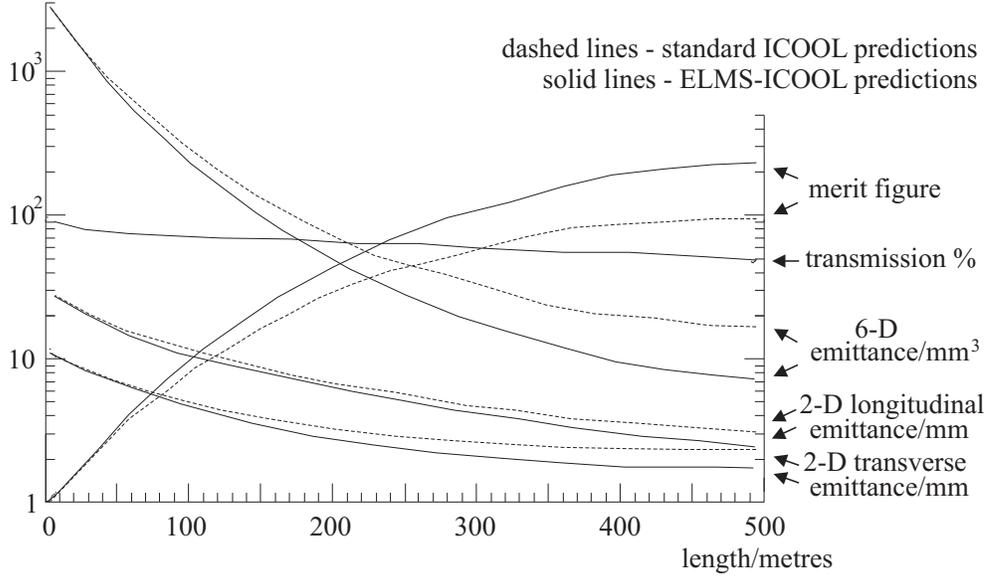}
\caption{Summary plot of the cooling performance with distance of an RFOFO ring.  }\label{f:rfofores}
\end{figure*}

The cooling in 6-D phase space has been studied in the context of the RFOFO cooling ring~\cite{RFOFO}. This offers a 
way to achieve simultaneous longitudinal and transverse cooling by using wedge shaped absorbers in a circular ring 
of absorbers and RF cavities. The results are shown in Fig.~\ref{f:rfofores} for standard ICOOL and ELMS-ICOOL. The 
cooling in transverse and normalised longitudinal emittances are combined to give the 6-D emittance. This may be 
used with the transmission efficiency to give an overall Merit Factor, defined as the increase in central 6-D phase 
space density including losses in transmission. While the transmission is the same (51\% \& 49\%) in 500m (15 
turns), ELMS predicts an increase in the Merit Factor to 230 from the value of 100 found by standard ICOOL. If the 
bunch charge is integrated over a finite central volume the cooling factor falls. However the improvement predicted 
by ELMS persists~\cite{SJH06}.

Instead of simulating the progressive cooling of muons towards an asymptotic normalised equilibrium emittance, the 
equilibrium emittance itself may be studied. Samples of 10$^5$ muons with $\beta _t=40$cm were considered with 
different values of incident transverse emittance. The value at which the normalised emittance following transport 
through 20cm of liquid H$_2$ did not change was noted. This equilibrium transverse emittance was found to be 1.60mm 
by standard ICOOL but 1.18mm by ELMS-ICOOL. 

\section{Conclusion}

We have pointed out that the cross section of muons on H$_2$ (with respect to energy loss, scattering and their 
correlation) can be accurately determined using fundamental principles and available atomic and molecular data. We 
have used this double differential cross section to derive a database of probability maps for thin absorbers. We 
have then tracked muons through thicker absorbers by Monte Carlo. Results are substantially more consistent with 
available experimental data than previous calculations. Simulations of a long MICE cooling channel and a ring 
cooler, and a study of equilibrium normalised emittance, consistently confirm that ionisation cooling in liquid 
hydrogen is significantly more effective than has been suggested by earlier simulations using less rigourous 
physics.

\end{document}